# Optimizing the dynamical behavior of a dual frequency parametric amplifier with quadratic and cubic nonlinearities

A. Dolev[a,1] and I. Bucher[a]

[a]*Dynamics Laboratory, Faculty of mechanical Engineering, Technion, Haifa, 3200003, Israel*

e-mail: amitdtechnion@gmail.com, ORCID: 0000-0003-1540-9784

e-mail: bucher@technion.ac.il, ORCID: 0000-0002-8815-4682

**Abstract**

The paper describes a novel parametric excitation scheme that acts as a tunable amplifier by controlling two pumping signals and two nonlinear feedback terms. By modulating the stiffness of a mechanical oscillator with a digital signal processor, low frequency inputs are projected onto a higher resonance frequency, thus exploiting the natural selective filtering of such structures. Described is an optimized dual-term nonlinear stiffness resonator that enhances the input signal level and the sensitivity to changes in both amplitude and phase, while limiting the obtained response to desired levels. This amplifier is geared to cases when the frequency of the input is known or measurable, like in rotating structures, while the amplitude and phase are too weak to be detected without amplification. It is shown that by tuning the cubic and quadratic feedback terms, the amplifier benefits from a nearly linear response behavior, while exploiting the benefits of nonlinear and pumping signal enhancements.

**Acknowledgments** The first author would like to acknowledge the generous financial support of the Israeli Ministry of Science, Technology & Space for the Applied science scholarship for engineering PhD students.

**The authors declare that there is no conflict of interest.**

# 1   Introduction

Parametric excitation (pumping) can be effective in signal amplification while squeezing noise [1], and can be highly frequency selective [2] .To achieve sufficient amplitudes, principal parametric resonance is commonly employed, and the pumped systems are operated close to the linear instability [3–6]. Nonlinear stiffness terms are considered to prevent excessive amplitude growth [7]. In order to amplify a harmonic non-resonant signal, additional pumping term is added to project its energy onto the resonance, thus amplifying the signal [8–13]. It has been shown that when only cubic nonlinearity is considered, the largest response levels appear at a close but different frequency than the ones exhibiting highest sensitivity [12]. By adding a quadratic nonlinear term, the response curve can be tuned, such that high sensitivity is achieved at close or identical frequency as principal parametric amplification.

---

[1] Corresponding author.

Single degree of freedom (SDOF) systems consisting nonlinear stiffness terms can exhibit a wide range of dynamical phenomena, as has already been reported by several researchers [7–10, 14–27]. Models describing continuous systems with multiple modes of vibration can still be adequately described as SDOF system with quadratic and cubic nonlinear terms, under specific operating conditions (e.g., Nayfeh and Lacabonara [14, 15]). The dynamic response of nonlinear systems to various external excitations was studied both numerically and analytically using various perturbation methods, [7, 16, 17] and their dynamical response to different parametric excitations was also studied [18–21] using diverse asymptotic methods. Additionally, the combined effect of external and parametric excitations was considered in prior studies [8–10, 12, 13, 22–27]. Nayfeh *et al* [28–30] extensively studied the influence of the various system parameters on the response of a SDOF system consisting quadratic and cubic nonlinear stiffness terms to different parametric excitations. In the majority of the works, the nonlinear terms have a physical reasoning, such as initial curvature leading to quadratic nonlinearity, and extension of the neutral axis leading to cubic nonlinearity [27, 31, 32]. In the present work, on the other hand, a linear system is considered and the nonlinearities are introduced in a controlled manner to achieve the desired amplifier dynamic behavior.

Amplifiers are widely used in different fields of engineering, and their role is to increase physical signals well above the noise floor, to achieve good signal to noise ratio. Because they are used to amplify various signals, (e.g., voltage, current, displacement, force) different amplification principles are utilized, and it has been shown that low loss mechanical amplifiers can outperform their electronic counterparts [33]. Moreover, linear amplifiers which are parametrically excited, also known as pumped, and operate near the principal parametric resonance [14] theoretically have infinite gain, because their response is unbounded by linear damping. In practice, the amplitude is bounded by nonlinear effects [18] or mechanical failures if large amplitudes evolve [34].

Although parametric amplifiers produce large amplitudes they are usually narrow banded, more than their equivalent linear resonators, therefore suitable for predetermined frequencies. To extend the parametric amplifier bandwidth a modified scheme was suggested by Dolev and Bucher [8–11]. According to the scheme, the amplifier is being pumped at two frequencies $(\omega_a, \omega_b)$, while the input signal to be amplified is the external force, whose frequency $\omega_r$ is assumed to be much lower than the natural frequency $\omega_n$. The various frequencies are algebraically related to each other according to $\omega_a = 2(\omega_b + \omega_r) \approx 2\omega_n$. The scheme utilizes the principal parametric resonance $(\omega_a \approx 2\omega_n)$ to produce large amplitudes and the combination resonance, $(\omega_b \approx \omega_n - \omega_r)$ to achieve sensitivity with respect to the external force. This scheme was verified numerically and validated experimentally for a SDOF system [9, 10], and a MDOF system [11, 12], when only cubic nonlinearity was considered. Furthermore, the scheme was used to project unbalance forces onto rotating system normal modes [12, 13] while it was rotated much slower than the critical speeds. This novel procedure allowed to balance the system without rotating it at dangerously high speeds. Indeed, large amplitudes were produced, however accurate tuning of the amplifier parameters to a fraction of 0.1Hz, was required to achieve satisfactory sensitivities. The tuning procedure is not trivial because the response is extremely narrow banded, especially when the damping is low. Accordingly, the parameters should by tuned with high precision, which required a lengthy and iterative process.

### 1.1 Improving the amplifier's performance

To increase the sensitivities and allow more flexibility in the parameters tuning process, it was suggested to add a quadratic nonlinearity to the feedback term, which allows to shape the backbone curve [28]. It is known that the cubic nonlinearity has either stiffening or softening effect on the frequency response, depending on its sign, while the quadratic nonlinearity has only a softening effect.

It was shown by Nayfeh [28], that by utilizing both nonlinearities, an equivalent nonlinearity may be considered. Thus, proper tuning of the nonlinear terms may lead to softening, hardening or near linear response, when small amplitudes are considered. The quadratic nonlinearity addition allows more flexibility in the selection of the amplifier parameters, which produce the desired dynamic response, i.e., large amplitude and high sensitivities. In what follows, it is shown that when the equivalent nonlinearity is tuned to zero, the amplification and sensitivities are proportional to each other, and may approach infinite values. In contrast, when the equivalent nonlinearity differs from zero or when only cubic stiffness is considered, the sensitivities have a finite maximal value, while the amplification is theoretically infinite.

In summary, the lumped parameters model of the considered parametric amplifier is depicted in Fig. 1, and its governing equation of motion is

$$m\ddot{y} + c\dot{y} + \left\{ \begin{pmatrix} 1 + \alpha_a \cos(\omega_a t + \varphi_a) \\ + \alpha_b \cos(\omega_b t + \varphi_b) \end{pmatrix} k_1 \\ + k_2 y + k_3 y^2 \right\} y = F\cos(\omega_r t + \varphi_r). \tag{1}$$

The amplifier is characterized by a point mass $m$, linear viscous damping $c$, linear stiffness $k_1$, quadratic stiffness $k_2$, cubic stiffness $k_3$, being pumped at frequencies $\omega_a$ and $\omega_b$ with the appropriate pumping magnitudes $\alpha_a$, $\alpha_b$ and corresponding phases $\varphi_a$, $\varphi_b$. The amplifier is fed by an external force with amplitude $F$, frequency $\omega_r$ and phase $\varphi_r$.

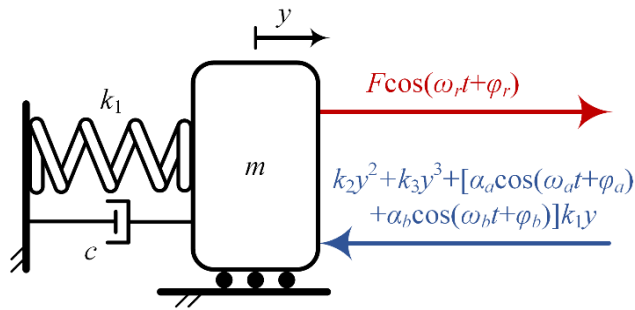

**Fig. 1** A SDOF parametric amplifier with quadratic and cubic controlled nonlinearities

The proposed scheme is designed to enhance the amplification, while keeping the sensitivity to the external force amplitude and phase, high. Most important is the ability to shift some of the external force energy at frequency $\omega_r$ to the designated natural frequency, $\omega_n$, when natural amplification takes place.

The paper begins with the analytical solution of the governing equation of motion. Afterwards, the linear system's stability is analyzed by seeking the transition hyperplane in the controlled parameter domain. The following section deals with the response sensitivity to the various parameters. The insights are then implemented to suggest a parameters tuning scheme for the desired optimal amplifier response. Next, the analytical solution is compared to numerically simulated results with good agreement, and lastly a summary and conclusions are provided.

## 2  Approximate analytical solution

In order to understand the role of the nonlinear terms and to analyze their effect, an asymptotic solution based on the multiple scales approximation is derived. To begin the analysis, first the governing equation of motion, Eq. (1), is transformed to the following form:

$$y'' + 2\hat{\zeta} y' + \left( \begin{pmatrix} 1 + \alpha_a \cos(\Omega_a \tau + \varphi_a) \\ + \alpha_b \cos(\Omega_b \tau + \varphi_b) \end{pmatrix} + \hat{\kappa}_2 y + \hat{\kappa}_3 y^2 \right) y = \hat{P} \cos(\Omega_r \tau + \varphi_r) \qquad (2)$$

$$\omega_n = \sqrt{k/m}, \quad \tau = \omega_n t, \quad \hat{\zeta} = \frac{c}{2m\omega_n}, \quad \hat{\kappa}_\bullet = \frac{k_\bullet}{m\omega_n^2}, \quad \Omega_\bullet = \frac{\omega_\bullet}{\omega_n}, \quad \hat{P} = \frac{F}{m\omega_n^2}.$$

Whereas, $\partial \bullet / \partial \tau \triangleq \bullet'$.

If the amplifier is lightly damped, a small dimensionless damping parameter can be defined as $\varepsilon \sim O(\hat{\zeta}) \ll 1$. Additionally, small amplitudes are assumed, therefore the following transformation is defined:

$$y = \varepsilon^2 x \,.^2 \qquad (3)$$

Eliminating $y$ from Eq.(2) via Eq.(3) and reordering it yields:

$$x'' + x = P \cos(\Omega_r \tau + \varphi_r) - \varepsilon^2 \kappa_3 x^3 - \\ \varepsilon \left( 2\zeta x' + \left( \gamma_a \cos(\Omega_a \tau + \varphi_a) + \gamma_b \cos(\Omega_b \tau + \varphi_b) \right) x + \kappa_2 x^2 \right) \qquad (4)$$

$$\hat{\zeta} = \varepsilon \zeta, \quad \alpha_\bullet = \varepsilon \gamma_\bullet, \quad \hat{P} = \varepsilon^2 P, \quad \varepsilon^2 \hat{\kappa}_2 = \varepsilon \kappa_2, \quad \varepsilon^4 \hat{\kappa}_3 = \varepsilon^2 \kappa_3.$$

---

[2] The latter transformation is chosen according to results which were previously achieved and measured in experiments [9].

A second order uniform solution is sought using the method of multiple scales [7, 17] in the form

$$x(\varepsilon,\tau) = x_0(\tau_0,\tau_1,\tau_2) + \varepsilon x_1(\tau_0,\tau_1,\tau_2) + \varepsilon^2 x_2(\tau_0,\tau_1,\tau_2). \tag{5}$$

Here, $\tau_0 = \tau$ is the fast times scale associated with changes occurring at frequency close to the natural frequency, now scaled to unity. The slow time scales, $\tau_1 = \varepsilon\tau$ and $\tau_2 = \varepsilon^2\tau$ are associated with amplitude and phase modulations due to the damping, nonlinearities, and resonances.

The first and second derivatives of $x(\varepsilon,\tau)$ with respect to $\tau$ are:

$$x' = D_0 x_0 + \varepsilon\left(D_1 x_0 + D_0 x_1\right) + \varepsilon^2\left(D_2 x_0 + D_1 x_1 + D_0 x_2\right) + O(\varepsilon^3) \tag{6}$$

$$x'' = D_0^2 x_0 + \varepsilon\left(2D_0 D_1 x_0 + D_0^2 x_1\right) + \\ \varepsilon^2\left(D_1^2 x_0 + 2D_0 D_2 x_0 + 2D_0 D_1 x_1 + D_0^2 x_2\right) + O(\varepsilon^3), \tag{7}$$

whereas, the notation $x_{i,\tau_j} \triangleq D_j x_i$ is adopted.

Eliminating $x$ and its temporal derivative from Eq.(4) via Eqns.(5)-(7), and collecting terms of the same powers of $\varepsilon$, one obtains:

$\underline{\varepsilon^0}:$
$$D_0^2 x_0 + x_0 = P\cos(\Omega_r \tau + \varphi_r) \tag{8}$$

$\underline{\varepsilon^1}:$
$$D_0^2 x_1 + x_1 = -2D_0 D_1 x_0 - 2\zeta D_0 x_0 - \gamma_a \cos(\Omega_a \tau + \varphi_a) x_0 \\ -\gamma_b \cos(\Omega_b \tau + \varphi_b) x_0 - \kappa_2 x_0^2 \tag{9}$$

$\underline{\varepsilon^2}:$
$$D_0^2 x_2 + x_2 = -D_1^2 x_0 - 2D_0 D_2 x_0 - 2D_0 D_1 x_1 - 2\zeta D_1 x_0 - 2\zeta D_0 x_1 \\ -\gamma_a \cos(\Omega_a \tau + \varphi_a) x_1 - \gamma_b \cos(\Omega_b \tau + \varphi_b) x_1 - 2\kappa_2 x_0 x_1 - \kappa_3 x_0^3 \tag{10}$$

The solution of Eq.(8), expressed in a complex form, is given by:

$$x_0 = A(\tau_1,\tau_2) e^{i\tau_0} + \Lambda e^{i(\Omega_r \tau_0 + \varphi_r)} + \text{cc}, \quad \Lambda = \frac{P}{2(1-\Omega_r^2)}. \tag{11}$$

Whereas, cc stands for the complex conjugate of the preceding terms. To update the solution up to the next order (i.e., $x_1$), Eq.(11) is substituted into Eq.(9). Any particular solution of the resulting equation contains secular terms and small-divisor terms when $\Omega_a \approx 2$, $\Omega_a \pm \Omega_r \approx 1$, $\Omega_b \approx 2$ and $\Omega_b \pm \Omega_r \approx 1$. According to the scheme [10] mentioned in Section 1 the pumping frequencies are tuned as $\Omega_a \approx 2$ and $\Omega_b + \Omega_r \approx 1$. Therefore, two detuning parameters $\sigma_a$ and $\sigma_b$ are introduced to convert the small-divisor terms into secular terms according to

$$\Omega_a = 2 + \varepsilon\sigma_a, \quad \Omega_b + \Omega_r = 1 + \varepsilon\sigma_b. \tag{12}$$

Once Eqns.(11) and (12) are substituted into Eq.(9), one can eliminate the terms producing secular terms in $x_1$, by enforcing:

$$-2iD_1A - 2i\zeta A - \frac{1}{2}\gamma_a \bar{A} e^{i(\sigma_a \tau_1 + \varphi_a)} - \frac{1}{2}\gamma_b \Lambda e^{i(\sigma_b \tau_1 + \varphi_b + \varphi_r)} = 0, \qquad (13)$$

whereas, $\bar{A}$ is the complex conjugate of $A$, and the particular solution of Eq.(9) becomes:

$$\begin{aligned}
x_1 &= \frac{2i\zeta \Lambda \Omega_r}{\Omega_r^2 - 1} e^{i(\Omega_r \tau_0 + \varphi_r)} + \frac{\gamma_a \Lambda}{2\left((\Omega_a - \Omega_r)^2 - 1\right)} e^{i((\Omega_a - \Omega_r)\tau_0 + \varphi_a - \varphi_r)} + \\
&\quad \frac{\gamma_a A}{2(2+\Omega_a)\Omega_a} e^{i((1+\Omega_a)\tau_0 + \varphi_a)} + \frac{1}{3}\kappa_2 A^2 e^{i2\tau_0} - \kappa_2 \left(\Lambda^2 + A\bar{A}\right) + \\
&\quad \frac{\gamma_a \Lambda}{2(\Omega_a + \Omega_r - 1)(\Omega_a + \Omega_r + 1)} e^{i((\Omega_a + \Omega_r)\tau_0 + \varphi_a + \varphi_r)} - \\
&\quad \frac{\gamma_b \Lambda}{2\left((\Omega_b - \Omega_r)^2 - 1\right)} e^{i((\Omega_b - \Omega_r)\tau_0 + \varphi_b - \varphi_r)} - \frac{\gamma_b A}{2(2+\Omega_b)\Omega_b} e^{i((1+\Omega_b)\tau_0 + \varphi_b)} + \\
&\quad \frac{\gamma_b A}{2(\Omega_b - 2)\Omega_b} e^{i((1-\Omega_b)\tau_0 - \varphi_b)} + \frac{\kappa_2 \Lambda^2}{4\Omega_r^2 - 1} e^{i(2\Omega_r \tau_0 + 2\varphi_r)} + \\
&\quad \frac{2\kappa_2 \Lambda A}{(\Omega_r - 2)\Omega_r} e^{i((1-\Omega_r)\tau_0 - \varphi_r)} + \frac{2\kappa_2 \Lambda A}{(2+\Omega_r)\Omega_r} e^{i((1+\Omega_r)\tau_0 + \varphi_r)} + \text{cc}
\end{aligned} \qquad (14)$$

To approximate the solution up to the second order, $x_2$, Eqns.(11), (12) and (14) are substituted into Eq.(10), while the terms $D_1 A$ and $D_1^2 A$ are eliminated via Eq.(13). Eliminating the secular terms in $x_2$ yields:

$$\begin{aligned}
&-\frac{\gamma_a \gamma_b \Lambda \left(7 + (\Omega_r - 4)\Omega_r\right) e^{i((\sigma_a - \sigma_b)\tau_1 + \varphi_a - \varphi_b - \varphi_r)}}{16(\Omega_r - 3)(\Omega_r - 1)} \\
&+ \frac{\gamma_b \Lambda \left(\sigma_b \left(\Omega_r^2 - 1\right) - i\zeta \left(\Omega_r (4 + \Omega_r) - 1\right)\right) e^{i(\sigma_b \tau_1 + \varphi_b + \varphi_r)}}{4\left(\Omega_r^2 - 1\right)} \\
&+ \left(\zeta^2 - \frac{3\gamma_a^2}{32} - \frac{\gamma_b^2}{2(\Omega_r - 3)(1+\Omega_r)} + 2\Lambda^2 \left(\frac{2\kappa_2^2 \left(6 - \Omega_r^2\right)}{4 - \Omega_r^2} - 3\kappa_3\right)\right) A \\
&+ \frac{1}{4}\gamma_a \sigma_a \bar{A} e^{i(\sigma_a \tau_1 + \varphi_a)} + \left(\frac{10\kappa_2^2}{3} - 3\kappa_3\right) A^2 \bar{A} - 2iD_2 A = 0.
\end{aligned} \qquad (15)$$

It can be shown that Eqns.(13) and (15) are the first two terms in a multiple scales analysis of:

$$2i(A' + \varepsilon\zeta A) + \frac{1}{4}\varepsilon\gamma_a \bar{A}(2-\varepsilon\sigma_a)e^{i(\sigma_a\varepsilon\tau+\varphi_a)}$$

$$+\frac{1}{4}\varepsilon\gamma_b\Lambda\left(2-\varepsilon\sigma_b + \varepsilon\frac{i\zeta(\Omega_r(4+\Omega_r)-1)}{\Omega_r^2 - 1}\right)e^{i(\sigma_b\varepsilon\tau+\varphi_b+\varphi_r)}$$

$$+\varepsilon^2\left(\begin{array}{l}\dfrac{\gamma_a\gamma_b\Lambda(7+(\Omega_r-4)\Omega_r)e^{i((\sigma_a-\sigma_b)\varepsilon\tau+\varphi_a-\varphi_b-\varphi_r)}}{16(\Omega_r-3)(\Omega_r-1)} + \\ \left(3\kappa_3 - \dfrac{10\kappa_2^2}{3}\right)A^2\bar{A} + \left(\begin{array}{l}\dfrac{3\gamma_a^2}{32} - \zeta^2 + \dfrac{\gamma_b^2}{2(\Omega_r-3)(1+\Omega_r)} + \\ 2\Lambda^2\left(3\kappa_3 - \dfrac{2\kappa_2^2(6-\Omega_r^2)}{4-\Omega_r^2}\right)\end{array}\right)A\end{array}\right) = 0. \quad (16)$$

Substituting the polar form:

$$A(\tau) = \frac{1}{2}a(\tau)e^{i\phi(\tau)}, \quad (17)$$

into Eq.(16) and separating into real and imaginary parts, yields:

$$\Re: \quad \begin{array}{l}\phi'a = \dfrac{1}{8}\varepsilon\left(\begin{array}{l}(2-\varepsilon\sigma_a)\gamma_a a\cos(\sigma_a\varepsilon\tau - 2\phi + \varphi_a) \\ +2(2-\varepsilon\sigma_b)\gamma_b\Lambda\cos(\sigma_b\varepsilon\tau - \phi + \varphi_b + \varphi_r) \\ -2\varepsilon\gamma_b\Lambda\zeta\left(\dfrac{(\Omega_r(4+\Omega_r)-1)}{\Omega_r^2-1}\right)\sin(\sigma_b\varepsilon\tau - \phi + \varphi_b + \varphi_r)\end{array}\right) \\ +\varepsilon^2\dfrac{\gamma_a\gamma_b\Lambda(7+(\Omega_r-4)\Omega_r)}{16(\Omega_r-3)(\Omega_r-1)}\cos((\sigma_a-\sigma_b)\varepsilon\tau - \phi + \varphi_a - \varphi_b - \varphi_r) \\ +\varepsilon^2\left(\begin{array}{l}\left(\dfrac{3\gamma_a^2}{64} - \dfrac{\zeta^2}{2} + \dfrac{\gamma_b^2}{4(\Omega_r-3)(1+\Omega_r)} + \Lambda^2\left(3\kappa_3 - \dfrac{2\kappa_2^2(6-\Omega_r^2)}{4-\Omega_r^2}\right)\right)a \\ +\left(\dfrac{3}{8}\kappa_3 - \dfrac{5\kappa_2^2}{12}\right)a^3\end{array}\right)\end{array} \quad (18)$$

$$\Im: \quad \begin{array}{l}a' = -\varepsilon\zeta a - \dfrac{1}{8}\varepsilon\left(\begin{array}{l}(2-\varepsilon\sigma_a)\gamma_a a\sin(\sigma_a\varepsilon\tau - 2\phi + \varphi_a) \\ +2(2-\varepsilon\sigma_b)\gamma_b\Lambda\sin(\sigma_b\varepsilon\tau - \phi + \varphi_b + \varphi_r) \\ +2\varepsilon\zeta\gamma_b\Lambda\left(\dfrac{(\Omega_r(4+\Omega_r)-1)}{\Omega_r^2-1}\right)\cos(\sigma_b\varepsilon\tau - \phi + \varphi_b + \varphi_r)\end{array}\right) \\ -\varepsilon^2\dfrac{\gamma_a\gamma_b\Lambda(7+(\Omega_r-4)\Omega_r)}{16(\Omega_r-3)(\Omega_r-1)}\sin((\sigma_a-\sigma_b)\varepsilon\tau - \phi + \varphi_a - \varphi_b - \varphi_r)\end{array} \quad (19)$$

To transform Eqns.(18)-(19) to an autonomous system, the following functions are defined:

$$\psi_a = \sigma_a \varepsilon \tau - 2\phi, \quad \psi_b = \sigma_b \varepsilon \tau - \phi, \quad \psi_{ab} = (\sigma_a - \sigma_b)\varepsilon \tau - \phi. \tag{20}$$

Differentiating Eq.(20) with respect to $\tau$ yields:

$$\psi'_a = \sigma_a \varepsilon - 2\phi', \quad \psi'_b = \sigma_b \varepsilon - \phi', \quad \psi'_{ab} = (\sigma_a - \sigma_b)\varepsilon - \phi'. \tag{21}$$

Seeking the steady-state solution, thus $\psi'_a = \psi'_b = \psi'_{ab} = 0$, and from Eq.(21):

$$\phi' = \frac{1}{2}\sigma_a \varepsilon = \sigma_b \varepsilon = (\sigma_a - \sigma_b)\varepsilon, \tag{22}$$

therefore, the following is defined:

$$\sigma = \sigma_b = \frac{1}{2}\sigma_a, \quad \psi = \psi_b = \psi_{ab} = \frac{1}{2}\psi_a. \tag{23}$$

Substituting Eqns.(20) and (23) into Eqns.(18)-(19), and solving for the steady-state, one sets $a' = \psi' = 0$, and the equations reduces to:

$$\Re: \quad 0 = \varepsilon \sigma a_0 - \frac{1}{4}\varepsilon \begin{pmatrix} (1-\varepsilon\sigma)\gamma_a a_0 \cos(2\psi_0 + \varphi_a) \\ +(2-\varepsilon\sigma)\gamma_b \Lambda \cos(\psi_0 + \varphi_b + \varphi_r) \\ -\varepsilon\gamma_b \Lambda \zeta \left( \dfrac{\Omega_r(4+\Omega_r)-1}{\Omega_r^2 - 1} \right) \sin(\psi_0 + \varphi_b + \varphi_r) \end{pmatrix}$$

$$-\varepsilon^2 \frac{\gamma_a \gamma_b \Lambda \left(7+(\Omega_r - 4)\Omega_r\right)}{16(\Omega_r - 3)(\Omega_r - 1)} \cos(\psi_0 + \varphi_a - \varphi_b - \varphi_r) \tag{24}$$

$$-\varepsilon^2 \begin{pmatrix} \left( \dfrac{3\gamma_a^2}{64} - \dfrac{\zeta^2}{2} + \dfrac{\gamma_b^2}{4(\Omega_r - 3)(1+\Omega_r)} + \Lambda^2 \left( 3\kappa_3 - \dfrac{2\kappa_2^2(6-\Omega_r^2)}{4-\Omega_r^2} \right) \right) a_0 \\ +\left( \dfrac{3}{8}\kappa_3 - \dfrac{5\kappa_2^2}{12} \right) a_0^3 \end{pmatrix}$$

$$\Im: \quad 0 = -\varepsilon \zeta a_0 - \frac{1}{4}\varepsilon \begin{pmatrix} (1-\varepsilon\sigma)\gamma_a a_0 \sin(2\psi_0 + \varphi_a) \\ +(2-\varepsilon\sigma)\gamma_b \Lambda \sin(\psi_0 + \varphi_b + \varphi_r) \\ +\varepsilon\zeta\gamma_b \Lambda \left( \dfrac{\Omega_r(4+\Omega_r)-1}{\Omega_r^2 - 1} \right) \cos(\psi_0 + \varphi_b + \varphi_r) \end{pmatrix} \tag{25}$$

$$-\varepsilon^2 \frac{\gamma_a \gamma_b \Lambda \left(7+(\Omega_r - 4)\Omega_r\right)}{16(\Omega_r - 3)(\Omega_r - 1)} \sin(\psi_0 + \varphi_a - \varphi_b - \varphi_r).$$

Whereas, the subscript 0 denotes the steady-state solution.

The steady-state amplitude, $a_0$, as a function of the steady-state phase, $\psi_0$, can be obtained from Eq.(25):

$$a_0 = -\gamma_b \Lambda \begin{pmatrix} 4\varepsilon\zeta\left(3+\Omega_r\left(\Omega_r^2+\Omega_r-13\right)\right)\cos\left(\psi_0+\varphi_b+\varphi_r\right) \\ +(1+\Omega_r)\begin{pmatrix} \gamma_a\varepsilon\left(7+(\Omega_r-4)\Omega_r\right)\sin\left(\psi_0+\varphi_a-\varphi_b-\varphi_r\right) \\ -4(\varepsilon\sigma-2)(\Omega_r-3)(\Omega_r-1)\sin\left(\psi_0+\varphi_b+\varphi_r\right) \end{pmatrix} \end{pmatrix} \quad (26)$$

$$\times \left\{ 4(\Omega_r-3)(\Omega_r-1)(\Omega_r+1)\left(4\zeta+\gamma_a(1-\varepsilon\sigma)\sin(2\psi_0+\varphi_a)\right)\right\}^{-1}.$$

Eliminating $a_0$ from Eq.(24) via Eq.(26), the steady state phase can be computed from a nonlinear transcendental equation. The latter can be cast in an efficient numerical form, by converting the transcendental equation to a tenth order polynomial using a similar procedure described in Appendix A in [10].

$$\sum_{n=0}^{5} \Delta_{2n} z^{2n} = 0, \quad z = e^{i\psi_0}, \quad \Delta_{2n} \in \mathbb{C}. \quad (27)$$

Once $a_0$ and $\psi_0$ are obtained, according to Eq.(5) the steady-state solution is:

$$\begin{aligned} x &= a_0 \cos\left(\frac{\Omega_a}{2}\tau-\psi_0\right)+2\Lambda\cos(\Omega_r\tau+\varphi_r)+\varepsilon x_1 + O(\varepsilon^2) \approx \\ &\approx a_0 \cos\left(\frac{\Omega_a}{2}\tau-\psi_0\right)+2\Lambda\cos(\Omega_r\tau+\varphi_r) \\ &+\varepsilon\left(\frac{4\zeta\Lambda\Omega_r}{1-\Omega_r^2}\sin(\Omega_r\tau+\varphi_r)+\frac{\gamma_b a_0}{2(\Omega_b-2)\Omega_b}\cos(\Omega_r\tau-\varphi_b-\psi_0)\right)+\text{AOT}= \\ &= a_0 \cos\left(\frac{\Omega_a}{2}\tau-\psi_0\right)+a_r\cos(\Omega_r\tau-\psi_r)+\text{AOT}, \end{aligned} \quad (28)$$

whereas, AOT stands for All Other Terms.

### 2.1 Discussion

The steady-state approximate response of the nonlinear, parametrically excited amplifier was obtained analytically up to the second order, when two harmonic terms, $\Omega_a/2$ and $\Omega_r$, are of interest. To this approximation, the quadratic and cubic nonlinear terms have a combined effect as can be seen in Eq.(16), and an equivalent nonlinearity can be defined as

$$\kappa_e = 3\kappa_3 - \frac{10}{3}\kappa_2^2. \quad (29)$$

When $\kappa_e > 0$, the nonlinearity is of the hardening type, and when $\kappa_e < 0$ it is of the softening type [35]. By tuning the cubic term as $\kappa_3 = 10/9\kappa_2^2$, the effective stiffness is nullified and the system behaves linearly up to the order of $\varepsilon^2$. It is worth noting, that the nonlinear terms also influence the response via the term

$2\Lambda^2\left(3\kappa_3 - 2\kappa_2^2\left(6-\Omega_r^2\right)/\left(4-\Omega_r^2\right)\right)A$ in Eq.(16). However, in Section 4 it is shown that this term has a negligible effect on the response.

By solving the Eq.(27), it was found that at steady-state the system may have up to five different solutions at the same frequency. A stable trivial solution for which $a_0 = 0$ does not exist due to the external force. The steady-state solutions stability is determined in a standard procedure [7, 28] by determining the behavior of the linear solutions of Eq.(16). Furthermore, a maximum of three stable solutions may exist at a given frequency.

Because the system acts as an amplifier, it is important to understand when large amplitudes can be produced. Therefore, one can define the criterion and a threshold value, $\gamma_{th}$.

$$4\zeta = \gamma_{th} \leq \gamma_a, \qquad (30)$$

for which the denominator of Eq.(26) can become null. In addition to the amplification capability, the amplifier sensitivities to the input signal amplitude and phase are important. The improved sensitivities may be downgraded or even lost if the principal parametric resonance overly dominates the response. Accordingly, the pumping magnitude should be set close to the threshold.

In a previous work, it was observed that even when $\gamma_a < \gamma_{th}$ large amplitudes were produced. A possible reason is that the second pumping frequency $(\Omega_b)$, having a different pumping amplitude, shifts the linear stability boundary as it feeds power into the system. It is important to find this boundary, since efficient operation is obtained while working in its vicinity. In the following section, the exact linear stability boundary is found for the case of two pumping frequencies and external forcing.

## 3     Stability regions of the linear system

The stable and unstable regions of the linear system are separated by hyper-surfaces that depend on the externally governed parameters. In order to compute these separating surfaces when the frequencies are tuned according to Eq.(12), the method of multiple scales is employed once more. Following a similar procedure to the one described in Section 2, one should analyze the following, which is identical to Eq.(16) without the nonlinear terms $(\kappa_2, \kappa_3)$:

$$2i(A' + \varepsilon\zeta A) + \frac{1}{2}\varepsilon\gamma_a \bar{A}(1-\varepsilon\sigma)e^{i(2\sigma\varepsilon\tau+\varphi_a)}$$

$$+\frac{1}{4}\varepsilon\gamma_b \Lambda\left(2-\varepsilon\sigma + \varepsilon\frac{i\zeta(\Omega_r(4+\Omega_r)-1)}{\Omega_r^2 - 1}\right)e^{i(\sigma\varepsilon\tau+\varphi_b+\varphi_r)} \qquad (31)$$

$$+\varepsilon^2\left(\begin{array}{c}\dfrac{\gamma_a\gamma_b\Lambda(7+(\Omega_r-4)\Omega_r)e^{i(\sigma\varepsilon\tau+\varphi_a-\varphi_b-\varphi_r)}}{16(\Omega_r-3)(\Omega_r-1)} + \\ +\left(\dfrac{3\gamma_a^2}{32} - \zeta^2 + \dfrac{\gamma_b^2}{2(\Omega_r-3)(1+\Omega_r)}\right)A\end{array}\right) = 0.$$

To obtain the solution of Eq.(31) one can write

$$A(\tau) = (B_r(\tau) + iB_i(\tau))e^{i\sigma\tau}, \qquad (32)$$

and separate Eq.(31) into real and imaginary parts:

$$\Re: \quad B_i' = \left[\frac{1}{4}\varepsilon(\gamma_a(1-\varepsilon\sigma)\sin(\varphi_a) - 4\zeta)\right]B_i$$

$$+\left[\frac{1}{64}\varepsilon\left(\begin{array}{c}3\gamma_a^2\varepsilon + 16\left(\dfrac{\gamma_b^2\varepsilon}{(\Omega_r-3)(\Omega_r+1)} - 2\varepsilon\zeta^2 - 4\sigma\right) + \\ 16\gamma_a(1-\varepsilon\sigma)\cos(\varphi_a)\end{array}\right)\right]B_r$$

$$+\left[\begin{array}{c}\dfrac{\gamma_b\varepsilon\Lambda}{32(\Omega_r-3)(\Omega_r-1)(\Omega_r+1)} \\ \left(\begin{array}{c}\gamma_a\varepsilon(\Omega_r+1)(7+(\Omega_r-4)\Omega_r)\cos(\varphi_a-\varphi_b-\varphi_r) \\ +4(3-\Omega_r)(\varepsilon\sigma-2)(\Omega_r^2-1)\cos(\varphi_b+\varphi_r) \\ +4\varepsilon\zeta(3-\Omega_r)(\Omega_r(4+\Omega_r)-1)\sin(\varphi_b+\varphi_r)\end{array}\right)\end{array}\right] \qquad (33)$$

$$\Im: \quad B_r' = \left[-\frac{1}{64}\varepsilon\left(\begin{array}{c}3\gamma_a^2\varepsilon + 16\left(\dfrac{\gamma_b^2\varepsilon}{(\Omega_r-3)(\Omega_r+1)} - 2\varepsilon\zeta^2 - 4\sigma\right) - \\ 16\gamma_a(1-\varepsilon\sigma)\cos(\varphi_a)\end{array}\right)\right]B_i$$

$$+\left[-\frac{1}{4}\varepsilon(\gamma_a(1-\varepsilon\sigma)\sin(\varphi_a) + 4\zeta)\right]B_r$$

$$-\left[\begin{array}{c}\dfrac{\gamma_b\varepsilon\Lambda}{32(\Omega_r-3)(\Omega_r-1)(\Omega_r+1)} \\ \left(\begin{array}{c}\gamma_a\varepsilon(\Omega_r+1)(7+(\Omega_r-4)\Omega_r)\sin(\varphi_a-\varphi_b-\varphi_r) \\ +4(3-\Omega_r)(\varepsilon\sigma-2)(\Omega_r^2-1)\sin(\varphi_b+\varphi_r) \\ +4\varepsilon\zeta(3+\Omega_r(\Omega_r+\Omega_r^2-13))\cos(\varphi_b+\varphi_r)\end{array}\right)\end{array}\right] \qquad (34)$$

Equations (33)-(34) can be written as the following differential equations:

$$\left\{\begin{matrix} B'_i \\ B'_r \end{matrix}\right\} = \begin{pmatrix} a & b \\ c & d \end{pmatrix} \left\{\begin{matrix} B_i \\ B_r \end{matrix}\right\} + \left\{\begin{matrix} e \\ f \end{matrix}\right\}, \qquad (35)$$

whose solution can be expressed as:

$$B_r = b_{r_1} e^{\beta_1 \tau} + b_{r_2} e^{\beta_2 \tau} + b_1, \quad B_i = b_{i_1} e^{\beta_1 \tau} + b_{i_2} e^{\beta_2 \tau} + b_2. \qquad (36)$$

Here, the $b_\bullet$ terms are constants and $\beta_1$ and $\beta_2$ are the eigenvalues, such that

$$\beta_1 \beta_2 = ad - bc, \quad \beta_1 + \beta_2 = a + d. \qquad (37)$$

The solution's stability is determined by the real part of the eigenvalues, $\beta_\bullet$. To find the stability transition hyper-surface, it is sufficient that one of the eigenvalues equals zero while the real part of the other one is non-positive. If one of the eigenvalues is set to zero, the other one must be real and negative, because $\{a,b,c,d\} \in \mathbb{R}$ and $a+d = -2\varepsilon\zeta$. Therefore, to compute the transition surfaces it is sufficient to check when does the determinant becomes null. The determinant can be written as a second order polynomial in $\sigma$:

$$F_2(\varepsilon, \gamma_a, \Omega_r)\sigma^2 + F_1(\varepsilon, \gamma_a, \Omega_r, \gamma_b, \zeta)\sigma + F_0(\varepsilon, \gamma_a, \Omega_r, \gamma_b, \zeta) = 0. \qquad (38)$$

The functions $F_\bullet$ and the stability transition hyper-surface are provided in Appendix A, Eqns. (44) and (45).

Choosing typical values for the parameters $\Omega_r = 0.1$, $\zeta = 2$ and $\varepsilon = 0.01$ (i.e., $\hat{\zeta} = 2\%$), one can plot the computed transition surface as is shown in Fig. 2(a)-(c); Fig. 2(d) was computed for $\Omega_r = 0.3$. Observing Fig. 2, one can clearly see that the stability depends on the detuning parameter $\sigma$, and both pumping magnitudes $\gamma_a$ and $\gamma_b$. The damping related parameter $\zeta$ and the external force frequency $\Omega_r$ also influence the stability transition surfaces. For lower values of damping, lower pumping magnitudes are required for the system to become unstable, in a similar manner to its effect on the Mathieu equation (e.g., [7]). The external force frequency, $\Omega_r$, affects the ratio between $\gamma_a$ and $\gamma_b$ leading to instability, as can be seen by comparing Fig. 2(c) and (d). When $\Omega_r$ is increased, $\Omega_b$ is reduced according to Eq.(12), therefore less power is pumped into the system via the term $\gamma_b \cos(\Omega_b t + \varphi_b) x$. Therefore, when $\Omega_r$ is increased, larger values of $\gamma_a$ are required for the system to become unstable for a given value of $\gamma_b$. Additionally, from Eq.(38) one can witness that the phases do not influence the stability.

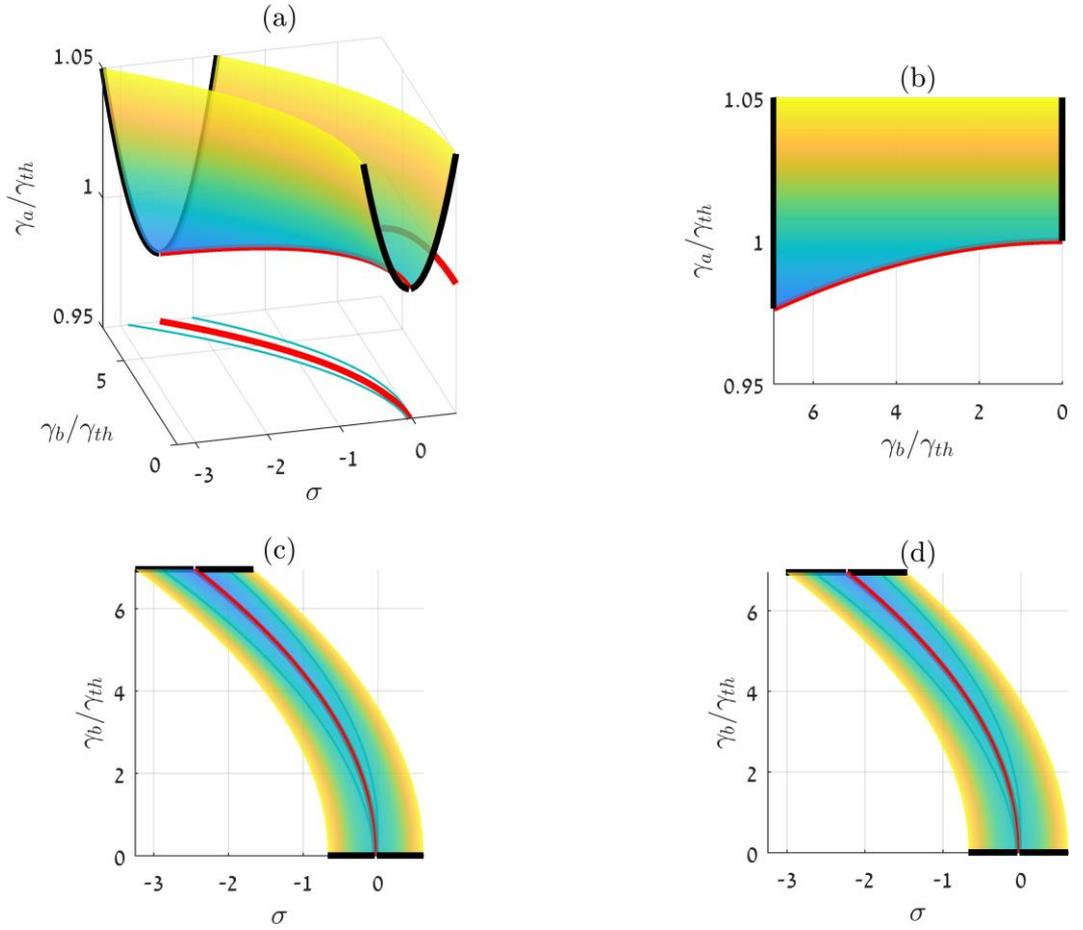

**Fig. 2** Stability transition hyper-surfaces – above the surfaces the system is unstable, while below it is stable, and the curve depicts the crest's centerline. (a)-(c) shows different projections of the same surface computed for $\Omega_r = 0.1, \zeta = 2, \varepsilon = 0.1$, and (d) shows a surface computed for $\Omega_r = 0.3, \zeta = 2, \varepsilon = 0.1$.

It is clear from Fig. 2(b) that the maximum value of $\gamma_a$ leading to instability is $\gamma_{th}$, and as $\gamma_b$ is increased, lower values of $\gamma_a$ are required. Moreover, the minimum value of $\gamma_a$ as a function of $\gamma_b$ leading to instability is depicted in Fig. 2 by the curve at the center of the crest, and it is calculated from Eq.(45) in Appendix A, by equating $\sigma_1 = \sigma_2$. This curve is given by:

$$G_6(\varepsilon, \Omega_r)\gamma_a^6 + G_4(\varepsilon, \Omega_r, \gamma_b, \zeta)\gamma_a^4 + G_2(\varepsilon, \Omega_r, \gamma_b, \zeta)\gamma_a^2 + G_0(\Omega_r, \zeta) = 0. \quad (39)$$

The functions $G_\bullet$ are also provided in Appendix A, Eq.(46), and the solution of Eq.(39) yields a single real valued curve, which is omitted for brevity. This is the improved criterion for selecting the pumping magnitudes, and is favored over Eq.(30).

# 4 Parameter sensitivity analysis

The underlying assumption is that the external force frequency $\Omega_r$ is known, but its amplitude and phase are not. It is assumed that the external force is weak and possibly buried in wideband noise, therefore accurate identification of its amplitude and phase requires selective amplification. To achieve the desired amplified response, proper tuning of the amplifier parameters needs to be carried out. Therefore, to better understand the influence of each parameter on the amplitude $a_0$, and sensitivities $\partial a_0 / \partial P$ and $\partial a_0 / \partial \varphi_r$ are required. A closed form expression for the approximate amplitude $a_0$ was derived in Section 2, and closed form expressions for the amplifier sensitivities are provided in Appendix B.

For the analysis, the following non-dimensional parameters were chosen:

$$\zeta = 2, \quad \varepsilon = 0.01, \quad \kappa_3 = 4, \quad \kappa_e = 5, \quad \gamma_b = 2\gamma_{th}, \quad (40)$$
$$\Omega_r = 0.1, \quad P = 1, \quad \varphi_a = \varphi_b = \varphi_r = 0.$$

According to Eq.(29), $\kappa_2 = 3/\sqrt{5}$, and the system is thus chosen to be marginally stable according to the linear system stability criterion, hence: $\gamma_a \approx 0.998\gamma_{th}$ and $\sigma \approx -0.2307$. Moreover, for the chosen parameters the system has only one stable static position at the origin.

The amplitudes $a_0$, $a_r$ and the sensitivities vs. the detuning parameter $\sigma$ are shown in Fig. 3, continuous lines represent stable solutions, dashed lines unstable solutions, and the dot marks the chosen detuning parameter, $\sigma \approx -0.2307$. From Fig. 3 (a)-(b) it is noticeable that the chosen pumping magnitudes produced parametric resonance, and large amplitudes were produced near the natural frequency. Still, at the external force frequency the amplitude remained relatively low. In addition, one can notice from Fig. 3(c)-(d) that for the chosen detuning parameter, the sensitivities are relatively high, although for lower $\sigma$, better sensitivities can be achieved. However, if lower $\sigma$ is used, the response amplitude decreases drastically as shown in Fig. 3(a).

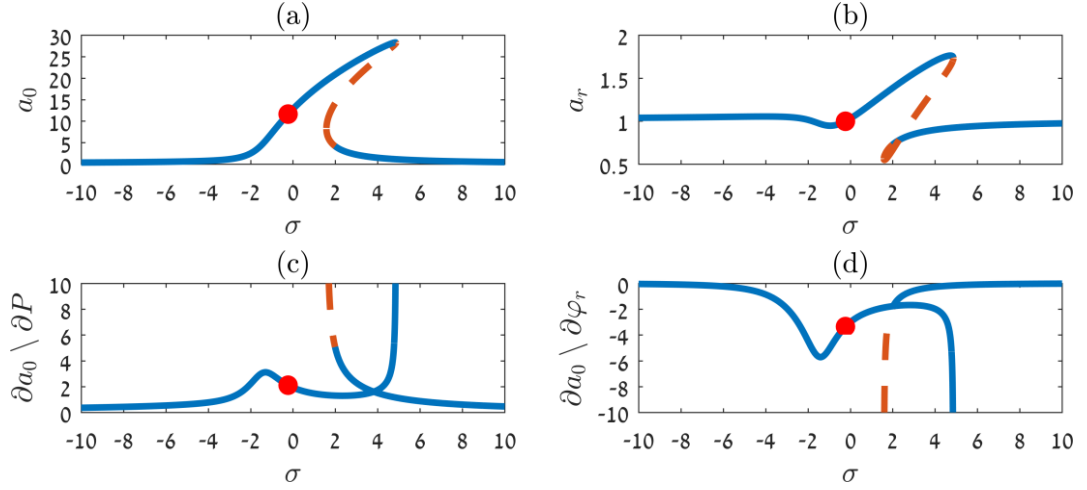

**Fig. 3** (a) The amplitude of the harmonic term near the natural frequency, $a_0$, and (b) at the external force frequency, $a_r$ vs. the detuning parameter. (c) The sensitivity with respect to the external force amplitude, $P$ and (d) with respect to the external phase, $\varphi_r$. Continuous lines represent stable solutions while dashed lines represent unstable solutions, and the dots mark the chosen detuning parameter.

### 4.1 Sensitivity to the nonlinear terms

The nonlinear terms in Eq.(40) were arbitrarily chosen within an acceptable range. To tune the nonlinear terms effectively, their influence on the response is studied. The amplitudes near the natural frequency, at the external force frequency and the sensitivities were computed as a function of $\kappa_2$ and $\kappa_3$ for $\sigma \approx -0.2307$ and are shown in Fig. 4. In the white region in each of the subfigures the effective stiffness $\kappa_e$ is negative, which means that the system is softening. This behavior is not desired therefore, disregarded in the analysis. Observing Fig. 4(a) which depicts the amplitude near the natural frequency vs. the nonlinear terms, one can whiteness that as the effective stiffness approaches zeros (i.e., the transition line between the colored and the white regions) the amplitude grows rapidly to infinity. The highlighted curves are not contour lines, but equal effective stiffness lines according to Eq.(29). Their minor deviation from the contour lines is probably due to the term $2\Lambda^2\left(3\kappa_3 - 2\kappa_2^2\left(6-\Omega_r^2\right)/\left(4-\Omega_r^2\right)\right)A$ in Eq.(16). Because the deviation is very small, in practice the latter term can be neglected, and only the effective stiffness should be considered. Observing Fig. 4(b)-(d), one can whiteness a similar phenomenon; as the effective stiffness decreases the amplitude at the external force frequency increases (b), the sensitivity to the external force amplitude increases (c) and the sensitivity to the external force phase absolute value increases.

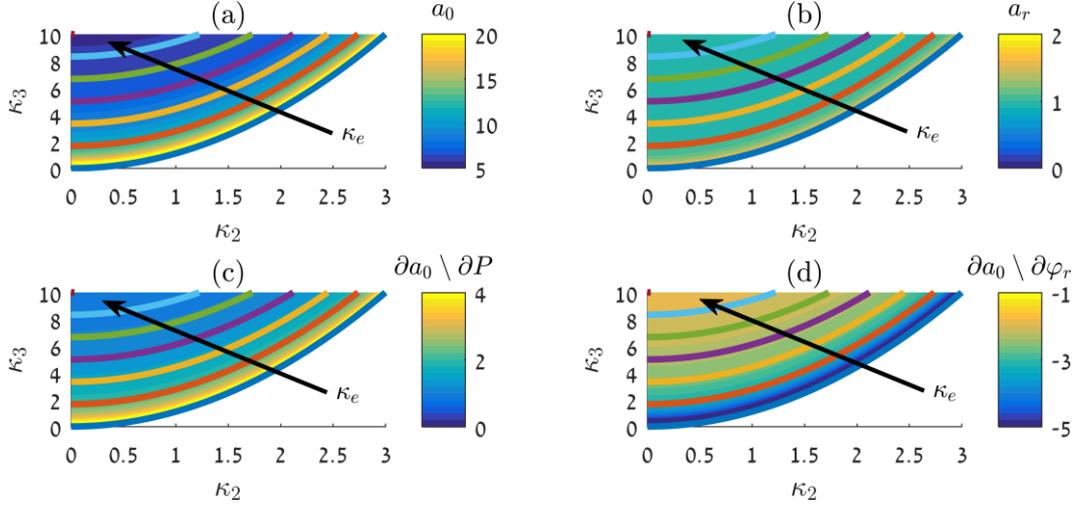

**Fig. 4** (a) The amplitude of the harmonic term near the natural frequency, $a_0$, and (b) at the external force frequency, $a_r$ vs. the quadratic and cubic stiffness terms. (c) The sensitivity with respect to the external force amplitude, $P$ and (d) with respect to the external phase, $\varphi_r$. The highlighted curves are equal effective stiffness lines.

Hence, to achieve the best performances the effective stiffness should be tuned to zero, $\kappa_e = 0$. One should note that the data shown in Fig. 4 was computed for $\sigma \approx -0.2307$, for which infinite amplitudes are produced if $\kappa_e \approx 0$. This dynamic behavior cannot be captured by the suggested model, as it suitable only for small amplitudes. Nevertheless, the system should be tuned as suggested with a different detuning parameter, say $\sigma \approx -1.1922$ that produces finite amplitudes, as shown in Fig. 5. The amplifier performances shown in Fig. 5 are superior to those shown in Fig. 3 in terms of amplification and sensitivities.

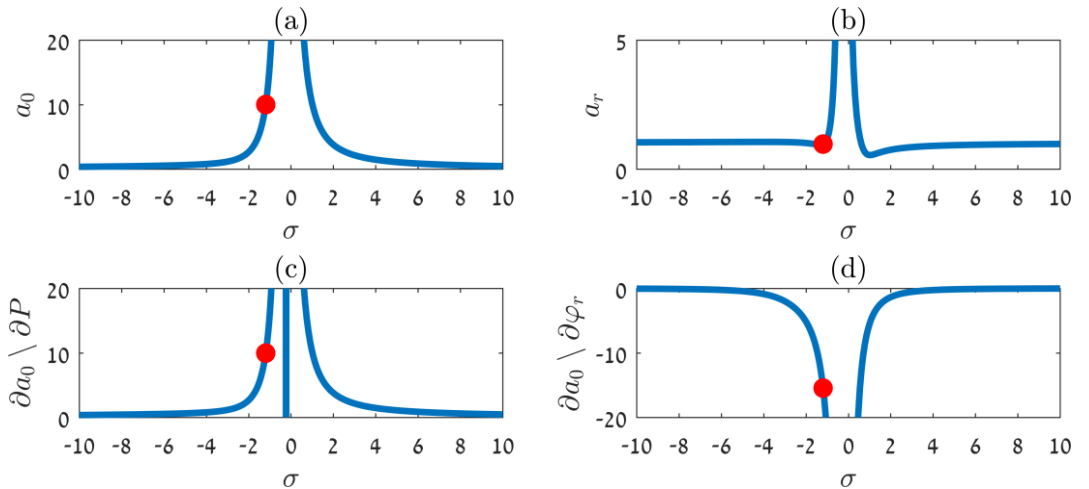

**Fig. 5** (a) The amplitude of the harmonic term near the natural frequency, $a_0$, and (b) at the external force frequency, $a_r$ vs. the detuning parameter. (c) The sensitivity with respect to the external force amplitude, $P$ and (d) with respect to the external phase, $\varphi_r$. The dots mark $\sigma \approx -1.1922$.

### 4.2  Sensitivity to the pumping magnitudes

The pumping magnitudes have a strong influence on the amplifier performance, especially when they are tuned in a way that leads to instability of the linear system, as discussed in Section 3. To understand their influence on the response, the amplitudes and sensitivities vs. $\gamma_a/\gamma_{th}$ and $\gamma_b/\gamma_{th}$ are shown in Fig. 6, here the parameters are set according to Eq.(40), $\kappa_2 = 3/\sqrt{5}$ and $\sigma = 0$. The continues lines in the subfigures represent the relation between $\gamma_a$ and $\gamma_b$ according to Eq.(39). Observing Fig. 6(a), as the pumping magnitudes are increased so as the amplitude near the natural frequency. For the amplitude at the external force frequency, shown in Fig. 6(b), the response is more complex; the amplitude $a_r$ does not necessarily grow when the pumping magnitudes are increased. There is a crest whose minimum values are highlighted by a white dashed line. For pumping values larger than these defined by the dashed line, the observation is the same as for the amplitude $a_0$. The reason for the latter observation is not trivial, because the amplitude $a_r$ depends directly on $\gamma_b$ and indirectly on $\gamma_a$ and $\gamma_b$ via $a_0$ and $\psi_0$, as can be seen in Eq.(28). The sensitivities shown in Fig. 6(c)-(d) behave similarly, and their highest absolute value is heighted by a dashed line.

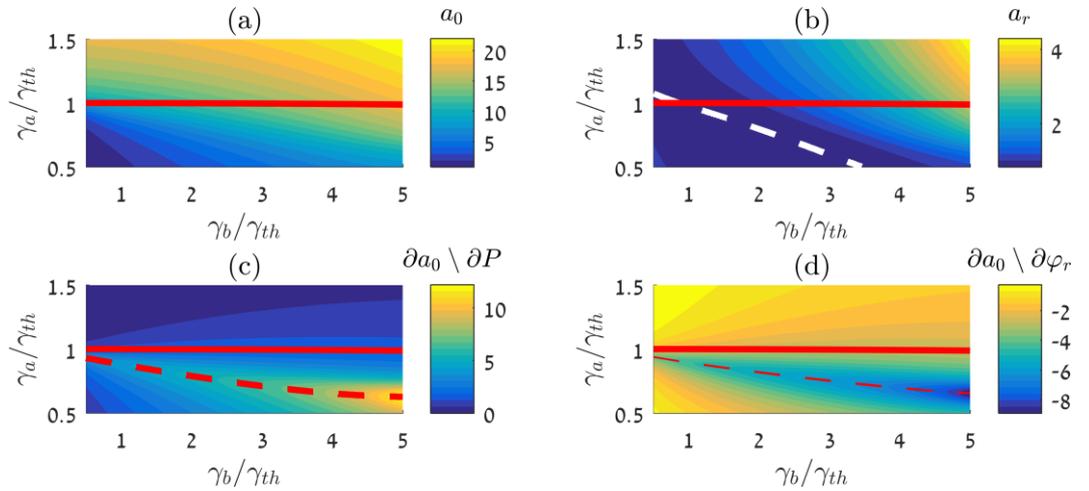

**Fig. 6** (a) The amplitude of the harmonic term near the natural frequency, $a_0$, and (b) at the external force frequency, $a_r$ vs. the normalized pumping magnitudes. (c) The sensitivity with respect to the external force amplitude, $P$ and (d) with respect to the external phase, $\varphi_r$. The continues lines represent the ratio between the pumping magnitudes according to Eq.(39), and the dashed lines trace the maximal (c) and minimal (b),(d) values.

Observing Fig. 6(a), (c)-(d), it seems that a compromise between the amplitude and sensitivities should be made when choosing the pumping magnitudes. However, in Section 4.1, it proved beneficial to tune the nonlinearities such that $\kappa_e = 0$ and set the detuning parameter as $\sigma \approx -1.1922$. The responses for this case, when $\kappa_3 = 10/9$ and $\kappa_2 = 1$ is shown in Fig. 7 for a wider range of pumping magnitudes. By comparing Fig. 6 to Fig. 7 it is clear that the dynamic responses differ. In the latter case, the amplitude $a_0$ shown in Fig. 7(a) attains infinite values as $\gamma_a$ is increased from zero for a given value of $\gamma_b$, if it is furtherly increased, the amplitude decreases. A similar phenomenon is obtained for $a_r$ and the sensitivities shown in Fig. 7(b)-(d). In contrast to the previous case, no compromise is made between the amplitude and the sensitivities when choosing the pumping magnitudes, hence tuning the effective stiffness to zero is beneficial.

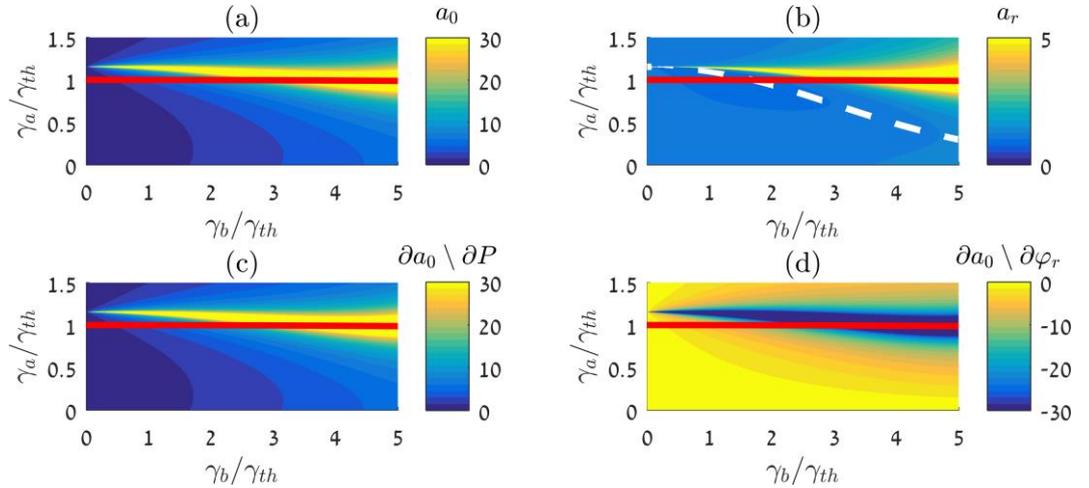

**Fig. 7** (a) The amplitude of the harmonic term near the natural frequency, $a_0$, and (b) at the external force frequency, $a_r$ vs. the normalized pumping magnitudes. (c) The sensitivity with respect to the external force amplitude, $P$ and (d) with respect to the external phase, $\varphi_r$. The continues lines represent the ratio between the pumping magnitudes according to Eq.(39), and the dashed line traced the minimal values (b).

To further understand how to tune the parameters $\gamma_a$, $\gamma_b$ and $\sigma$ the dynamics along the curve given by Eq.(39) vs. $\gamma_b$ and $\sigma$ is shown for $\kappa_3 = 10/9$ and $\kappa_2 = 1$ in Fig. 8. Along the curve given by Eq.(39) $\gamma_a/\gamma_{th} \approx 1$ for the chosen $\gamma_b$ values as depicted in Fig. 2(b). To help the reader, on Fig. 8(a) three level line (5, 10 and 15) are highlighted by dashed white lines. It can be seen that as $\gamma_b$ is increased the gradient in $\sigma$ direction decreases and lower detuning levels are needed to achieve the same amplitude. Similar behavior is observed for the sensitivities Fig. 8(c)-(d) and the amplitude at the external force Fig. 8(b). A section view of Fig. 8 for $\gamma_b/\gamma_{th} = 2$ was previously shown in Fig. 5.

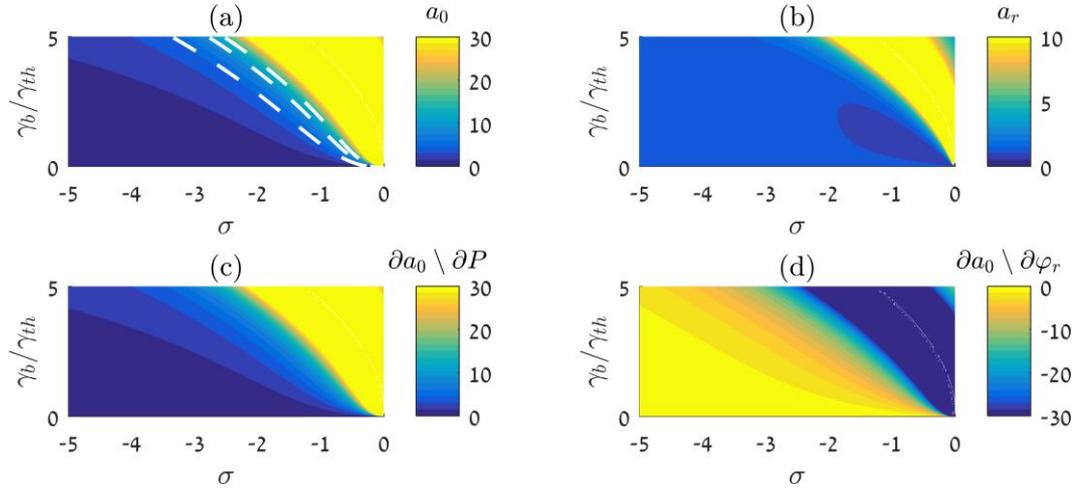

**Fig. 8** (a) The amplitude of the harmonic term near the natural frequency, $a_0$, and (b) at the external force frequency, $a_r$ vs. $\sigma$ and $\gamma_b/\gamma_{th}$ along the curve defined by Eq.(39). (c) The sensitivity with respect to the external force amplitude, $P$ and (d) with respect to the external phase, $\varphi_r$. The dashed white lines are the level lines, 5,10 and 15.

In conclusion, it was found that if the equivalent stiffness is set to zero, no comprise between the amplitude and sensitivities is made when selecting the pumping magnitudes and detuning parameter. To tune the pumping parameters it is suggested to first select $\gamma_b$ and then tune $\gamma_a$ according to Eq.(39), and lastly select the detuning parameter $\sigma$ according to the desired response, as shown in Fig. 5.

### 4.3  Sensitivity to the external force and pumping phases

The suggested system is designed to oscillate at a frequency close to its natural frequency while the input signal is at a much lower frequency. Additionally, it was shown that variations in the external force amplitude or phase, leads to variations in the amplitude $a_0$. To identify the input's amplitude and phase, a single measurement of $a_0$ is not enough. Interestingly, the pumping phase $\varphi_b$ always appears with $\varphi_r$ as a sum in the governing equations (e.g., Eq.(15)), therefore their influence on the response is the same, as shown in Fig. 9(a). Because $\varphi_b$ is controlled, $\varphi_r$ can be found in a $\pi$ width interval. Furthermore, one can devise a method to obtain $\varphi_r$ by tuning $\varphi_b$ and find either the maxima or minima, as done by Tresser *et al* [12].

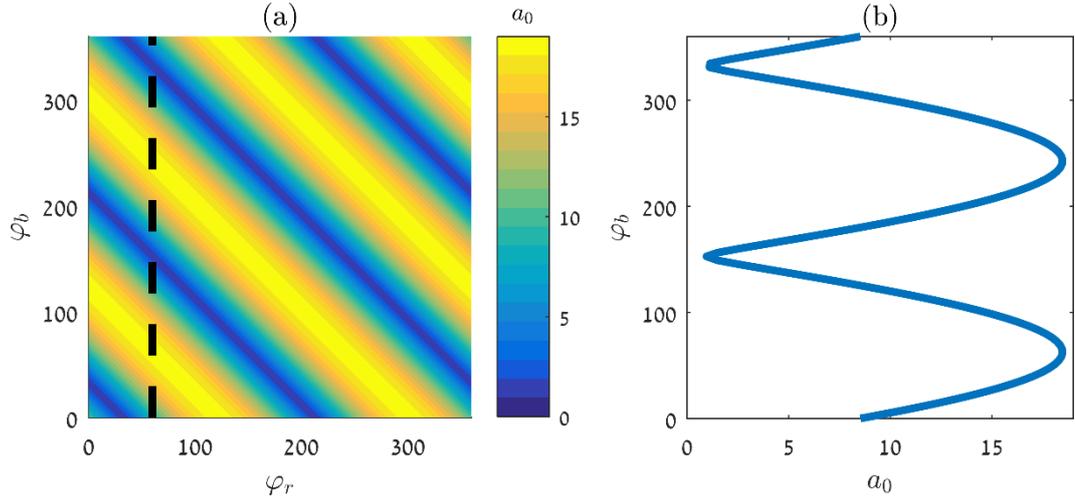

**Fig. 9** (a) The amplitude near the natural frequency, $a_0$ vs. the external force phase $\varphi_r$ and the pumping phase $\varphi_b$. A section view is shown for $\varphi_r = 60°$ (b), defined by the dashed line in (a).

The influence of $\varphi_a$ and the sum $\varphi_b + \varphi_r$ on the dynamics is shown in Fig. 10. The response is cyclic with respect to either $\varphi_a$ or $\varphi_b + \varphi_r$, but with a different periods, either $2\pi$ or $\pi$ in accordance. Moreover, $\varphi_a$ is completely controllable, unlike the phase sum, hence it can be used to modify the amplifier dynamics.

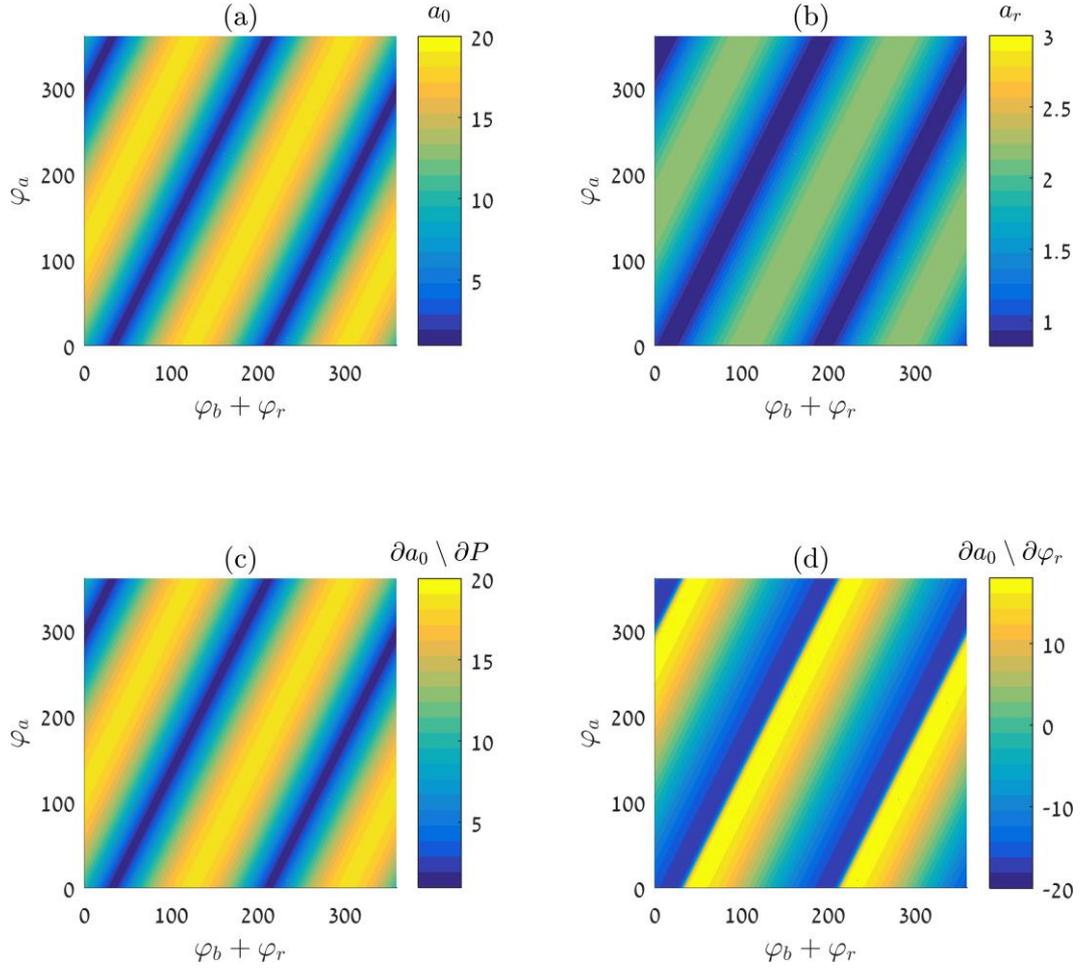

**Fig. 10** (a) The amplitude of the harmonic term near the natural frequency, $a_0$, and (b) at the external force frequency, $a_r$ vs. the pumping phase $\varphi_a$ and sum of the external force and other pumping phase $\varphi_b + \varphi_b$. (c) The sensitivity with respect to the external force amplitude, $P$ and (d) with respect to the external phase, $\varphi_r$ vs. the phases.

## 5    Parameters tuning for optimal response

The effect of various parameter was examined in Section 4 based on the analytical solution of the nonlinear and linear governing equation. As the studied system is designed to operate as an amplifier; one would like to tune it to produce the largest possible amplification and sensitivities. It was found that if the effective stiffness is set to zero, no compromise between the amplitude and sensitivity should be made when tuning the other parameters. To tune the effective stiffness to zero, according to Eq.(29), one should set $\kappa_3 = 10/9 \kappa_2^2$, and according to the analysis they should be of $O(1)$, hence they are set as $\kappa_3 = 10/9$ and $\kappa_2 = 1$. Next it is suggested to tune the pumping magnitudes according to Eq.(39), to ensure the ability to produce large amplitudes. To further simplify this procedure, a maximum allowable stiffness variation, $\Delta k_1$ can be defined as a design criterion, and it is related to the pumping magnitudes in the following manner:

$$\varepsilon(\gamma_a + \gamma_b) = \Delta k_1, \quad 0 < \Delta k_1 \leq 1 \tag{41}$$

By substituting Eq.(41) to Eq.(39), either $\gamma_a$ or $\gamma_b$ can be simply computed. Once $\kappa_2$, $\kappa_3$ and $\gamma_b$ are set, the detuning parameter $\sigma$ is tuned to produce the desired amplitude and sensitivity according to Fig. 8, as shown in Fig. 5. Finer adjustments can then be made by tuning $\varphi_a$, while $\varphi_b$ is used to obtain $\varphi_r$.

## 6 Verification via numerical simulations

Prior to considering the amplifier response to the excitations, the equilibrium points of the system are calculated from Eq.(4). Setting the excitations $\gamma_a$, $\gamma_b$, $P$ and all the derivatives to zero, the equilibrium points are

$$x_{eq} = 0, \frac{-\kappa_2 \pm \sqrt{\kappa_2^2 - 4\kappa_3}}{2\varepsilon\kappa_3} \tag{42}$$

In what follows, the system parameters were chosen according to an existing experimental rig [9] whose effective nonlinear stiffness is nullified:

$$\zeta = 2, \quad \varepsilon = 0.01, \quad \kappa_2 = 1, \quad \kappa_3 = 10/9. \tag{43}$$

For these chosen parameters, a single real equilibrium point exists at the origin, $x_{eq} = 0$. The pumping magnitude $\gamma_b$ was set as $\gamma_{th}$, $\gamma_a$ was set according to Eq.(39), and the phases were set to zero unless stated differently.

Frequency sweep results for $\Omega_r = 0.1$, $P = 1$ and two values of $\kappa_e$ are shown in Fig. 11; for both $\kappa_e$ values, a single equilibrium point exists at the origin. The amplitudes near the natural frequency and at the external force frequency when $\kappa_e = 0$ are shown in Fig. 11(a), (c) accordingly. The analytical results are depicted by continuous and dashed lines, while the simulations by hollow circular markers. It can be seen in Fig. 11(a), that the analytical results predict linear response, while the simulations indicate nonlinear behavior when relatively large amplitudes are produced. It is interesting to see that the amplitude $a_r$ decreases while $a_0$ grows, indicating that energy shifts from one frequency $(\Omega_r)$ to another $(\Omega_a/2)$. In Fig. 11(b), (d), the amplitudes are shown for the case when $\kappa_e = 5$. The effective stiffness has a major impact on the response, as it bends the curve to the right. Moreover, multiple solution regions exist, and one of the solutions is unstable (shown by a dashed line). Here, as for $\kappa_e = 0$, there is a good agreement between the analytical and numerical simulations, especially when the amplitudes are small.

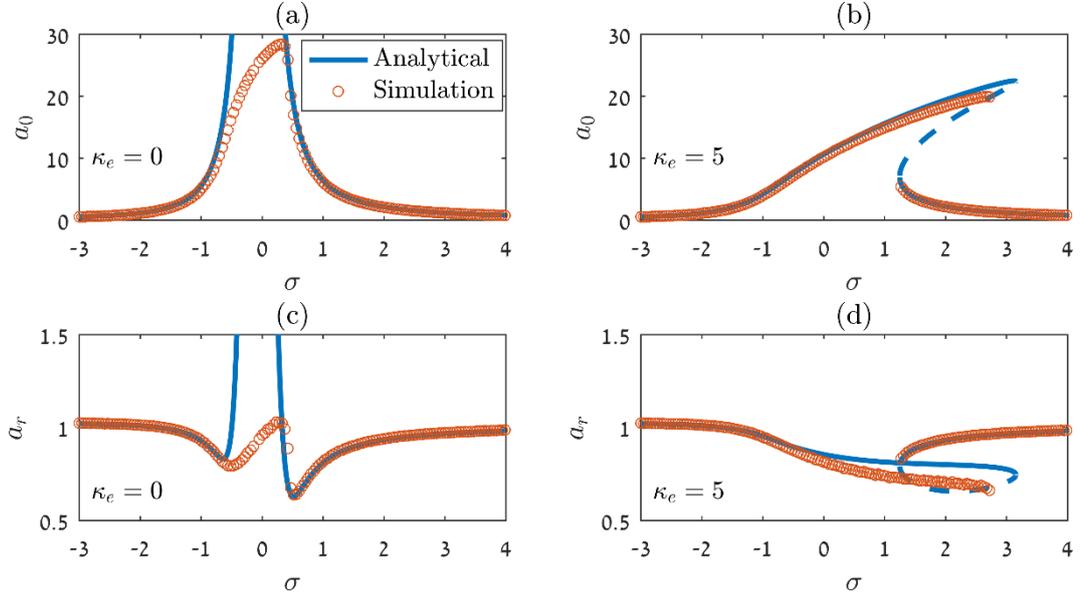

**Fig. 11** Analytically computed (continuous/ dashed lines) and numerically simulated (hollow circular markers) frequency sweep for two effective stiffness values. (a), (b) The amplitude of the harmonic term near the natural frequency, $a_0$, and (c), (d) at the external force frequency, $a_r$.

As mentioned throughout the manuscript, the sensitivity to the external force amplitude is very important, and also the ability to amplify a signal out of a wide frequency band. In Fig. 12 the amplitude $a_0$ and the sensitivity $\partial a_0 / \partial P$ are shown for $\Omega_r = 0.1, 0.3$. The detuning parameters were chosen to produce $a_0 = 8$ for $P = 1$, as was shown in Fig. 5. In Fig. 12(a), (b) the analytically computed amplitudes are shown by continuous lines, and the simulated by hollow circular markers. To compute the sensitivity from the simulated results, a smooth spline was fitted and is shown by a continuous line. In subfigures (c) and (d) the analytically and numerically computed sensitivities are shown by continuous lines. Again, there is a good agreement between the analytically and numerically computed amplitudes which starts to deviate as the amplitudes grow, and the agreement between sensitivities is in accordance. Additionally, the amplifier is suitable for signals in the frequency band $0.1 \leq \Omega_r \leq 0.3$. It can be furtherly shown that the amplifier is suitable for a wider frequency band, if neither super-harmonic nor subharmonic nor other combination resonances occur [7]. Nevertheless, sensitivity larger than unity is superior to the one obtained by a linear resonator, when used as an amplifier.

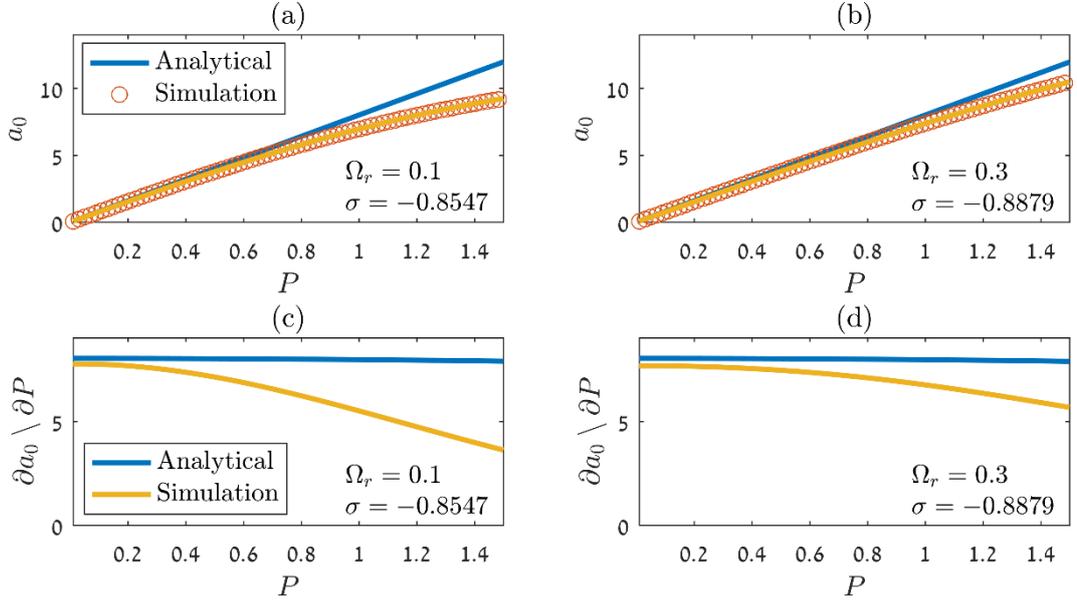

**Fig. 12** Analytically (continuous lines) and numerically (hollow circular markers) computed amplitudes and sensitivities for two external force frequencies. (a), (b) The amplitude of the harmonic term near the natural frequency, $a_0$, and (c), (d) the sensitivity, $\partial a_0 / \partial P$.

Other scenarios were studied as well, the response vs. phase variations. In Fig. 13, the response vs. the pumping phase $\varphi_a$ is shown for $P = 0.5$ and two external force frequencies $\Omega_r = 0.1, 0.3$, while the detuning parameters were set as in Fig. 12. It is noticeable that the analytical and numerical solution agree, and indeed the response is cyclic with a $2\pi$ period.

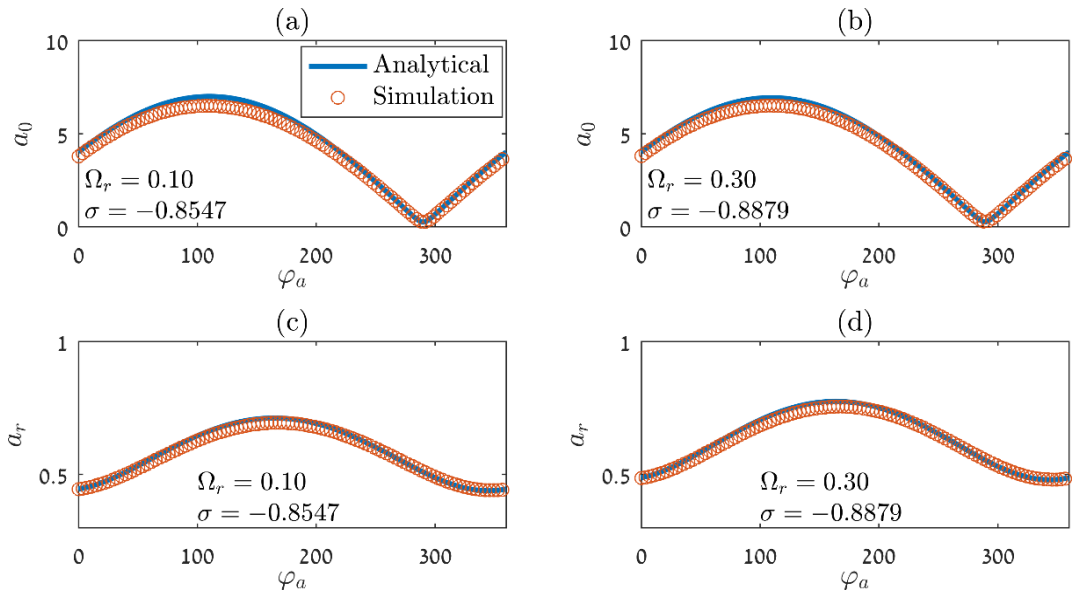

**Fig. 13** Analytically computed (continuous lines) and numerically simulated (hollow circular markers) responses vs. the pumping phase $\varphi_a$ for two external force frequencies. (a), (b) The amplitude of the harmonic term near the natural frequency, $a_0$, and (c), (d) at the external force frequency, $a_r$.

Fig. 14, depicts the response vs. the phase sum $\varphi_b + \varphi_r$ for $P$ and $\Omega_r$ as in the previous cases, while the detuning parameters were set to produce $a_0 = 10$ for $P = 1$. As in the previous cases, the analytical and numerical results agree, and the response is cyclic with a $\pi$ period.

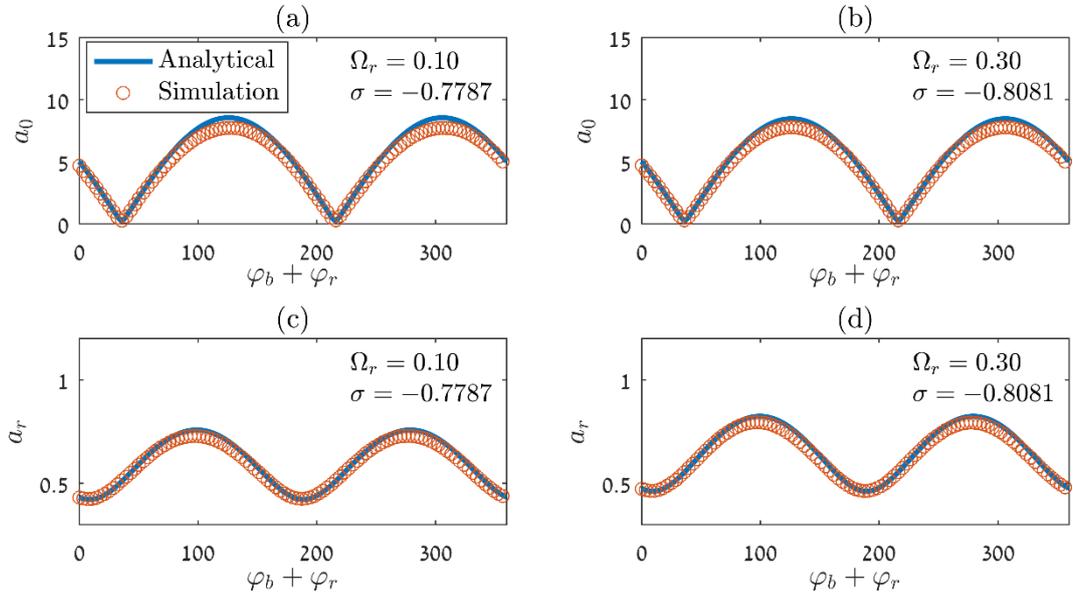

**Fig. 14** Analytically computed (continuous lines) and numerically simulated (hollow circular markers) responses vs. the phase sum $\varphi_b + \varphi_r$ for two external force frequencies. (a), (b) The amplitude of the harmonic term near the natural frequency, $a_0$, and (c), (d) at the external force frequency, $a_r$.

# 7 Conclusions

In this work, a parameter tuning methodology for a nonlinear SDOF dual frequency parametrically excited amplifier was derived. The topology of the proposed amplifier is unique and it allows to tune low frequency inputs by shifting the energy to a chosen natural frequency. The relevant parameters can be controlled by a digital signal processor that implements a feedback that determines the desired parameter levels, as was done in [9, 11, 12]. By implementing this parameter tuning scheme, which is described in Section 5, the amplifier was made highly sensitive to the input signal amplitude and phase, while large amplitudes, hence good signal to noise ratio, are produced. The latter was achieved by tuning the nonlinear terms such that the effective stiffness is nullified, yielding a linear dynamic behavior up to the second order, as shown in Fig. 11. This configuration produces relatively linear relationship between the external force amplitude $P$ at the frequency $\Omega_r$, and the response amplitude $a_0$ close to the natural frequency, $\Omega_a/2$, as depicted in Fig. 12. In addition, by setting the effective stiffness to zero, achieving improved amplification by tuning the pumping magnitudes and detuning parameter yielded improved sensitivities. In case $\kappa_e \neq 0$ the latter does not occur, and a compromise between the amplification and sensitivities needs to be made. The input signal phase can be detected within a $\pi$ interval by scanning the pumping frequency $\varphi_b$, because $\varphi_r$ is linearly related to it as shown in Fig. 9.

The derivation of the suggested scheme was made possible by analytically solving the linear and nonlinear governing equation, Eq.(4), using the method of multiple scales. Given the obtained steady-state solution of the nonlinear system, the sensitivities with respect to the external force amplitude and phase were derived, though omitted for brevity. The linear equation was solved to derive the transition hyper-surfaces between stable and unstable regions, which were then used to define a relation between the pumping magnitudes and detuning parameter ensuring the ability to produce large amplitudes.

Lastly, comparisons between the analytically and numerically obtained solutions were performed, with good agreement for various cases. The system parameters were set to match the properties of an existing experimental rig found in the lab. Future work will apply this approach to an experiment which is being constructed.

## Appendix A

For brevity, the following functions appearing in Eq.(38), are provided here:

$$F_2(\varepsilon,\gamma_a,\Omega_r) = -256(\gamma_a^2\varepsilon^2 - 16)(\Omega_r - 3)^2(\Omega_r + 1)^2$$

$$F_1(\varepsilon,\gamma_a,\Omega_r,\gamma_b,\zeta) = 128\varepsilon(\Omega_r - 3)(\Omega_r + 1)$$
$$\left((\Omega_r - 3)(\Omega_r + 1)(\gamma_a^2 + 32\zeta^2) - 16\gamma_b^2\right)$$

$$F_0(\varepsilon,\gamma_a,\Omega_r,\gamma_b,\zeta) = 9\gamma_a^2\varepsilon^2(\Omega_r - 3)^2(\Omega_r + 1)^2 + \tag{44}$$
$$32\gamma_a^2(\Omega_r - 3)(\Omega_r + 1)\left(3\gamma_b^2\varepsilon^2 - 2(4 + 3\varepsilon^2\zeta^2)(\Omega_r - 3)(\Omega_r + 1)\right) +$$
$$256\begin{pmatrix}\gamma_b^4\varepsilon^2 - 4\gamma_b^2\varepsilon^2\zeta^2(\Omega_r - 3)(\Omega_r + 1) + \\ 4\zeta^2(4 + \varepsilon^2\zeta^2)(\Omega_r - 3)^2(\Omega_r + 1)^2\end{pmatrix}.$$

By solving Eq.(38), the stability transition surfaces are computed, and given by:

$$\sigma_{1,2}(\gamma_a,\gamma_b,\Omega_r,\zeta,\varepsilon) = \pm\frac{1}{16(-16 + \gamma_a^2\varepsilon^2)(\Omega_r - 3)^2(\Omega_r + 1)^2}$$

$$\begin{pmatrix}\mp 4\varepsilon(\Omega_r - 3)(\Omega_r + 1)\left(-16\gamma_b^2 + (\gamma_a^2 + 32\zeta^2)(\Omega_r - 3)(\Omega_r + 1)\right) \\ +\sqrt{(\Omega_r - 3)^2(\Omega_r + 1)^2} \\ \sqrt{\begin{array}{l}(9\gamma_a^6\varepsilon^4 - 65536\zeta^2)(\Omega_r - 3)^2(\Omega_r + 1)^2 + \\ 96\gamma_a^4\varepsilon^2(\Omega_r - 3)(\Omega_r + 1) \\ (\gamma_b^2\varepsilon^2 - 2(2 + \varepsilon^2\zeta^2)(\Omega_r - 3)(\Omega_r + 1)) + \\ 256\gamma_a^2\begin{pmatrix}\gamma_b^4\varepsilon^4 - 4\gamma_b^2\varepsilon^2(2 + \varepsilon^2\zeta^2)(\Omega_r - 3)(\Omega_r + 1) + \\ 4(4 + 8\varepsilon^2\zeta^2 + \varepsilon^4\zeta^4)(\Omega_r - 3)^2(\Omega_r + 1)^2\end{pmatrix}\end{array}}\end{pmatrix} \tag{45}$$

For brevity, the following functions appearing in Eq.(39), are provided here:

$$G_6 = 9\varepsilon^4(\Omega_r - 3)^2(\Omega_r + 1)^2$$
$$G_4 = 96\varepsilon^2(\Omega_r - 3)(\Omega_r + 1)\left(\gamma_b^2\varepsilon^2 - 2(2 + \varepsilon^2\zeta^2)(\Omega_r - 3)(\Omega_r + 1)\right)$$
$$G_2 = 256\begin{pmatrix}\gamma_b^4\varepsilon^4 - 4\gamma_b^2\varepsilon^2(2 + \varepsilon^2\zeta^2)(\Omega_r - 3)(\Omega_r + 1) \\ +4(4 + 8\varepsilon^2\zeta^2 + \varepsilon^4\zeta^4)(\Omega_r - 3)^2(\Omega_r + 1)^2\end{pmatrix} \tag{46}$$
$$G_0 = -65536\zeta^2(\Omega_r - 3)^2(\Omega_r + 1)^2$$

## Appendix B

The amplifier sensitivities to variations in the input signal are key features. The amplifier is operated such that the response is dominated by the harmonic term at the frequency $\Omega_a/2 \approx 1$ with amplitude $a_0$, see Eq.(28). Therefore, the sensitivity to variations in the amplitude and phase of the input signal are studied via the sensitivity of the latter harmonic term. First, a closed form expression for the amplitude sensitivity is derived, and then in an analogous manner a closed form expression for the phase sensitivity is derived.

The sensitivity to the external force amplitude is computed as follows:

$$S_F^{y,a_0} = \frac{\partial \varepsilon^2 a_0}{\partial F} = \frac{1}{m\omega_n^2}\frac{\partial a_0}{\partial P}, \quad (47)$$

whereas, the term $\partial a_0 / \partial P$ is computed by taking the derivative of Eq.(26) with respect to $P$, which can be written as:

$$\frac{\partial a_0}{\partial P} = \Xi_1(\sigma,\varphi_a,\varphi_b,\gamma_a) + \Xi_2(\sigma,\varphi_a,\varphi_b,\gamma_a)\frac{\partial \psi_0}{\partial P}. \quad (48)$$

To compute $\partial a_0 / \partial P$ one must first compute $\partial \psi_0 / \partial P$, which can be computed from the nonlinear transcendental phase equation. Taking its derivative with respect to $P$, when only the numerator is considered, yields:

$$\frac{\partial \psi_0}{\partial P} = \Xi_3(\sigma,\varphi_a,\varphi_b,\gamma_a,\gamma_b,\kappa_2,\kappa_3). \quad (49)$$

Hence, the amplifier sensitivity to variations in the external force amplitude can be evaluated via:

$$\frac{\partial a_0}{\partial P} = \Xi_1(\sigma,\varphi_a,\varphi_b,\gamma_a) + \Xi_2(\sigma,\varphi_a,\varphi_b,\gamma_a)\Xi_3(\sigma,\varphi_a,\varphi_b,\gamma_a,\gamma_b,\kappa_2,\kappa_3). \quad (50)$$

The sensitivity to the external force phase, $\varphi_r$ is computed as follows:

$$S_{\varphi_r}^{y,a_0} = \frac{\partial \varepsilon^2 a_0}{\partial \varphi_r} = \varepsilon^2 \frac{\partial a_0}{\partial \varphi_r}. \quad (51)$$

Following a similar procedure, the sensitivity with respect to variations in the external force phase is obtained in a closed form:

$$\frac{\partial a_0}{\partial \varphi_r} = \Theta_1(\sigma,\varphi_a,\varphi_b,\gamma_a) + \Theta_2(\sigma,\varphi_a,\varphi_b,\gamma_a)\Theta_3(\sigma,\varphi_a,\varphi_b,\gamma_a,\gamma_b,\kappa_2,\kappa_3). \quad (52)$$

The functions $\Xi_\bullet$, $\Theta_\bullet$, $\bullet = 1,2,3$ are omitted for brevity, and are written only as functions of the relevant controlled parameters.

# References

1. Rugar, D., Grütter, P.: Mechanical parametric amplification and thermomechanical noise squeezing. Phys. Rev. Lett. 67, 699–702 (1991). doi:10.1103/PhysRevLett.67.699
2. Yu, M.-F., Wagner, G.J., Ruoff, R.S., Dyer, M.J.: Realization of parametric

resonances in a nanowire mechanical system with nanomanipulation inside a scanning electron microscope. Phys. Rev. B. 66, 73406 (2002). doi:10.1103/PhysRevB.66.073406

3. Krylov, S., Gerson, Y., Nachmias, T., Keren, U.: Excitation of large-amplitude parametric resonance by the mechanical stiffness modulation of a microstructure. J. Micromechanics Microengineering. 20, 15041 (2009). doi:10.1088/0960-1317/20/1/015041

4. Zalalutdinov, M., Olkhovets, a., Zehnder, a., Ilic, B., Czaplewski, D., Craighead, H.G., Parpia, J.M.: Optically pumped parametric amplification for micromechanical oscillators. Appl. Phys. Lett. 78, 3142–3144 (2001). doi:10.1063/1.1371248

5. Oropeza-Ramos, L.A., Burgner, C.B., Turner, K.L.: Robust micro-rate sensor actuated by parametric resonance. Sensors Actuators A Phys. 152, 80–87 (2009). doi:10.1016/j.sna.2009.03.010

6. Zaqarashvili, T. V., Oliver, R., Ballester, J.L.: Parametric Amplification of Magnetosonic Waves by an External, Transversal, Periodic Action. Astrophys. J. 569, 519–530 (2002). doi:10.1086/339288

7. Nayfeh, A.H., Mook, D.T.: Nonlinear oscillations. Wiley-VCH, Weinheim (2008)

8. Dolev, A., Bucher, I.: A parametric amplifier for weak, low-frequency forces. In: In ASME 2015 International Design Engineering Technical Conferences and Computers and Information in Engineering Conference. p. V006T10A046. ASME, Boston, Mass., USA (2015)

9. Dolev, A., Bucher, I.: Experimental and numerical validation of digital, electromechanical, parametrically excited amplifiers. J. Vib. 138, (2016). doi:10.1115/1.4033897

10. Dolev, A., Bucher, I.: Tuneable, non-degenerated, nonlinear, parametrically-excited amplifier. J. Sound Vib. 361, 176–189 (2016). doi:10.1016/j.jsv.2015.09.048

11. Dolev, A., Bucher, I.: Dual frequency parametric excitation of a nonlinear multi degree of freedom amplifier with a digitally modified topology. (2017)

12. Tresser, S., Dolev, A., Bucher, I.: Dynamic balancing of super-critical rotating structures using slow-speed data via parametric excitation. (2017)

13. Tresser, S., Bucher, I.: A method for balancing high-speed rotors using low

rotation speed measured data through parametric excitation. In: Proceedings of the 11th International conference Vibration in Rotating machinery (VIRM). , manchester (2016)

14. Ali, H.N., Walter, L.: On the discretization of spatially continuous systems with quadratic and cubic nonlinearities. JSME Int. J. Ser. C. 41, 510–531 (1998). doi:10.1299/jsmec.41.510

15. Nayfeh, A.H., Lacarbonara, W.: On the discretization of distributed-parameter systems with quadratic and cubic nonlinearities. Nonlinear Dyn. 13, 203–220 (1997). doi:10.1023/A:1008253901255

16. Nayfeh, A., Bouguerra, H.: Non-linear response of a fluid valve. Int. J. Non. Linear. Mech. 25, 433–449 (1990). doi:10.1016/0020-7462(90)90031-4

17. Nayfeh, A.H.: Perturbation Methods. Wiley-VCH, Weinheim (2008)

18. Rhoads, J.F., Shaw, S.W.: The impact of nonlinearity on degenerate parametric amplifiers Rhoads. Appl. Phys. Lett. 96, 234101 (2010). doi:10.1063/1.3446851

19. DeMartini, B.E., Rhoads, J.F., Turner, K.L., Shaw, S.W., Moehlis, J.: Linear and nonlinear tuning of parametrically excited MEMS oscillators. J. Microelectromechanical Syst. 16, 310–318 (2007). doi:10.1109/JMEMS.2007.892910

20. Rhoads, J.F., Shaw, S.W., Turner, K.L., Moehlis, J., DeMartini, B.E., Zhang, W.: Generalized parametric resonance in electrostatically actuated microelectromechanical oscillators. J. Sound Vib. 296, 797–829 (2006). doi:10.1016/j.jsv.2006.03.009

21. Maccari, a.: Multiple resonant or non-resonant parametric excitations for nonlinear oscillators. J. Sound Vib. 242, 855–866 (2001). doi:10.1006/jsvi.2000.3386

22. Troger, H., Hsu, C.: Response of a nonlinear system under combined parametric and forcing excitation. ASME Trans. Ser. E J. (1977)

23. Perkins, N.: Modal interactions in the non-linear response of elastic cables under parametric/external excitation. Int. J. Non. Linear. Mech. 27, 233–250 (1992). doi:10.1016/0020-7462(92)90083-J

24. Szabelski, K., Warminski, J.: Self-excited system vibrations with parametric and external excitations. J. Sound Vib. 187, 595–607 (1995). doi:10.1006/jsvi.1995.0547

25. Zhang, W., Tang, Y.: Global dynamics of the cable under combined parametrical and external excitations. Int. J. Non. Linear. Mech. 37, 505–526 (2002). doi:10.1016/S0020-7462(01)00026-9

26. El-Bassiouny, A., Abdelhafez, H.: Prediction of bifurcations for external and parametric excited one-degree-of-freedom system with quadratic, cubic and quartic non-linearities. Math. Comput. Simul. 57, 61–80 (2001). doi:10.1016/S0378-4754(01)00292-0

27. Kim, C.H., Lee, C.-W., Perkins, N.C.: Nonlinear vibration of sheet metal plates under interacting parametric and external excitation during manufacturing. In: In ASME 2003 International Design Engineering Technical Conferences and Computers and Information in Engineering Conference. pp. 2481–2489. ASME, Chicago, Ill., USA (2003)

28. Zavodney, L.D., Nayfeh, a. H., Sanchez, N.E.: The response of a single-degree-of-freedom system with quadratic and cubic non-linearities to a principal parametric resonance. J. Sound Vib. 129, 417–442 (1989). doi:10.1016/0022-460X(89)90433-1

29. Zavodney, L.D., Nayfeh, A.H.: The response of a single-degree-of-freedom system with quadratic and cubic non-linearities to a fundamental parametric resonance. J. Sound Vib. 120, 63–93 (1988). doi:10.1016/0022-460X(88)90335-5

30. Nayfeh, A.H.: The response of single degree of freedom systems with quadratic and cubic non-linearities to a subharmonic excitation. J. Sound Vib. 89, 457–470 (1983). doi:10.1016/0022-460X(83)90347-4

31. Chu, H.: Influence of large amplitudes on free flexural vibrations of rectangular elastic plates. J. Appl. Mechnics. (1956)

32. Marín, J., Perkins, N., Vorus, W.: Non-linear response of predeformed plates subject to harmonic in-plane edge loading. J. Sound Vib. 176, 515–529 (1994). doi:10.1006/jsvi.1994.1393

33. Bannon, F., Clark, J.: High-Q HF microelectromechanical filters. IEEE J. solid-state. 35, 512–526 (2000). doi:10.1109/4.839911

34. Billah, K.Y., Scanlan, R.H.: Resonance, Tacoma Narrows bridge failure, and undergraduate physics textbooks. Am. J. Phys. 59, 118–124 (1991). doi:10.1119/1.16590

35. Kovacic, I., Brennan, M.: The Duffing equation: nonlinear oscillators and

their behaviour. (2011)